\documentclass[onecolumn,superscriptaddress,amsmath,aps,prl]{revtex4}

\usepackage{graphicx}
\usepackage{wrapfig}
\usepackage{appendix}
\usepackage{subfigure}
\usepackage{textcomp}
\newcommand{\etal}{\emph{et al.}~}

\begin{document}

\title{Reducing the number of ancilla qubits and the gate count required for creating large controlled operations}
\author{Katherine L. Brown}
   \affiliation{Hearne Institute for Theoretical Physics and Department of Physics and Astronomy, Louisiana State University, Baton Rouge, LA, 70803, USA}
\author{Anmer Daskin}
   \affiliation{Department of Chemistry \& Computer Science, Purdue University, West Lafayette, Indiana, 47907, USA}
\author{Sabre Kais}
   \affiliation{Department of Chemistry \& Computer Science, Purdue University, West Lafayette, Indiana, 47907, USA}
\author{Jonathan P. Dowling }
   \affiliation{Hearne Institute for Theoretical Physics and Department of Physics and Astronomy, Louisiana State University, Baton Rouge, LA, 70803, USA}
   
\date{\today}
   
\begin{abstract}    
In this paper we show that it is possible to adapt a qudit scheme for creating a controlled-Toffoli created by Ralph \etal [Phys. Rev. A \textbf{75} 011213] to be applicable to qubits. While this scheme requires more gates than standard schemes for creating large controlled gates, we show that with simple adaptations it is directly equivalent to the standard scheme in the literature. This scheme is the most gate-efficient way of creating large controlled unitaries currently known, however it is expensive in terms of the number of ancilla qubits used. We go on to show that using a combination of these standard techniques presented by Barenco \etal [Phys. Rev. A \textbf{52} 3457 (1995)] we can create an n-qubit version of the Toffoli using less gates and the same number of ancilla qubits as recent work using computer optimization. This would be useful in any architecture of quantum computing where gates are cheap but qubit initialization is expensive. 
\end{abstract}

\maketitle
Making a unitary controlled on other qubits is an essential task for many algorithms in quantum computing \cite{Shor1997,Daskin2011,Wang2008}. In this paper we focus on a particular problem, making a unitary which is already controlled on one qubit, controlled on $n-1$ further qubits. These highly controlled unitaries (i.e.~unitaries controlled on more than one other qubit) are useful in numerous quantum algorithms including the oracle in the binary welded tree algorithm \cite{Childs2003} and quantum simulation \cite{Daskin2011,Wang2008}. Barenco \etal\cite{Barenco1995} outlined several techniques to make controlled unitaries in 1995, this work was expanded on in Nielsen and Chuang \cite{Nielsen2000} to provide a technique for making a n-qubit version of the CNOT gate using $14n-13$ operations and $n-2$ ancilla. Other work has explored this problem in the context of using computational algorithms to optimize circuit layout using the decomposition procedures proposed by Barenco \etal\cite{Barenco1995} and known commutation relations \cite{Maslov2003,Maslov2005,Scott2008,Wille2008,Miller2009,Miller2011}. However, while these techniques are useful if we have native controlled-square-root-not gates, in the case where the only native two qubit operation is a CNOT or C-Phase they perform worse than the technique in Nielsen and Chuang \cite{Nielsen2000} for the same number of ancilla. 

An interesting alternative technique was proposed by Ralph \etal\cite{Ralph2007} and implemented experimentally by Lanyon \etal\cite{Lanyon2008} who used additional levels in one of the subsystems of the controlled gate to reduce the overall number of operations required. In this work we will show that when converted to qubits this technique is directly equivalent to the one from Nielsen and Chuang \cite{Nielsen2000} using the same number of operations, and requiring the same number of ancilla. We go on to show that by using the techniques from Nielsen and Chuang \cite{Nielsen2000} combined with other decomposition procedures from Barenco \etal \cite{Barenco1995} it is possible to reduce the number of ancilla qubits required from $n-2$ to $2\sqrt{n-1}$ at the expense of double the number of operations. Our new techniques requires less operations for the same number of ancilla qubits when compared to existing decomposition schemes if we assume both schemes have the ability of perform a controlled square root of NOT gate \cite{Maslov2003,Maslov2005,Scott2008,Wille2008,Miller2009,Miller2011}.

For the rest of this work we will represent a CNOT gate controlled on n-qubits as C$^{n}$X, and a generally local unitary controlled on n qubits as C$^{n}$U. Given the ability to perform general local unitaries, and a CNOT gate we can make a Toffoli gate using nine local unitaries and six CNOT gates \cite{Shende2008}. However, if are prepared to accept an approximate Toffoli gate, we can use the Margolus gate, which is equivalent to a Toffoli gate and a controlled controlled phase. This procedure requires three CNOT gates and eight local unitaries \cite{Margolus1994,DiVincenzo1998}. An alternative set of decomposition procedures are used if we assume the ability to perform the CV gate in a single operation. Here a standard Toffoli uses two CNOT gates and three CV gates \cite{Barenco1995,Miller2011}. A more efficient decomposition is the Peres gate \cite{Peres1985,Miller2009}, which is the equivalent of a Toffoli gate with an additional CNOT. This decomposition uses only one CNOT gate, and three CV gates. In this work we can use the more efficient decomposition procedures because our Toffoli gates are arranged symmetrically with no other operations in between them. 

\begin{figure}
\includegraphics[width=12cm]{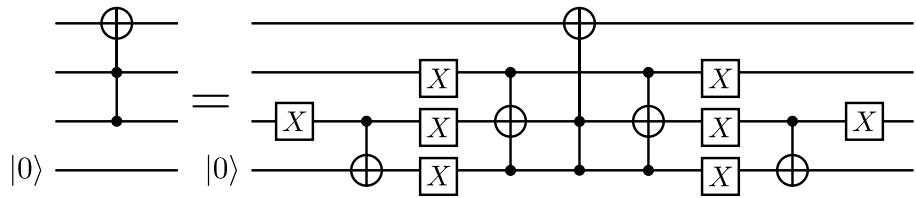}
\caption{A qubit equivalent of the Lanyon Toffoli, a single Toffoli gate is created but requires the use of three Toffoli gates. All lines represent qubits.}
\label{BToffoli}
\end{figure}

Ralph \emph{et al.}~\cite{Ralph2007} and Lanyon \emph{et al.~}\cite{Lanyon2008} demonstrate that by using a qutrit they can generate a Toffoli gate using only three CNOT gates, two standard Pauli X operations, and two implementations of a three-level version of the Pauli X operation. One requirement of Lanyon \emph{et al.}~is that the CNOT gates act trivially on the $|2\rangle$ level of the qutrit. Replacing the qutrit with two qubits would require each CNOT gate to be replaced with a Toffoli gate as shown in fig \ref{BToffoli}.  A qubit version of the Lanyon Toffoli gate is therefore impossible since we would need to consume three Toffoli gates to build a single Toffoli gate. 

\begin{figure}
\includegraphics[width=12.9cm]{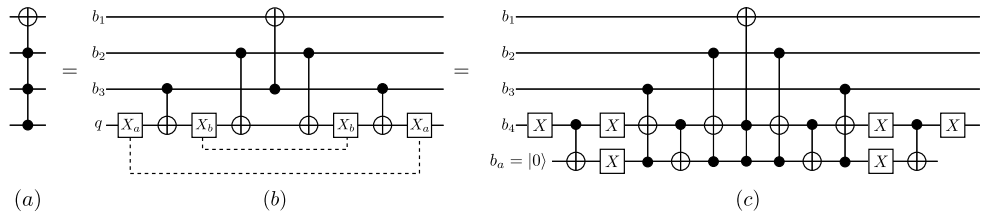}
\caption{A circuit for performing a C$^{3}$X gate  \textbf{(a)} using three qubits and a ququart where the ququart system is labelled \emph{q}. $X_{a}$ switches $|0\rangle$ and $|2\rangle$, $X_{b}$ switches $|1\rangle$ and $|3 \rangle$ and the CNOT gates act trivially on levels $|2\rangle$ and $|3\rangle$ of the ququart. \textbf{(b)} using five qubits.}
\label{Lquart}
\end{figure}

When we consider replacing the qudit with $d=4$ (ququart), needed by Lanyon \emph{et al.}~\cite{Lanyon2008} to generate a C$^{3}$U gate, we use five Toffoli gates to generate a controlled Toffoli gate. In fig \ref{Lquart} we show the original circuit proposed by Lanyon \emph{et al.}~and our qubit adaptation. Breaking down our Toffoli gates into CNOT gates and general local unitaries means we require 53 operations. The standard decomposition by Barenco \emph{et al.}~\cite{Barenco1995} requires 44 operations. Therefore further simplification is needed. 

\begin{figure}[h]
\includegraphics[width=12.9cm]{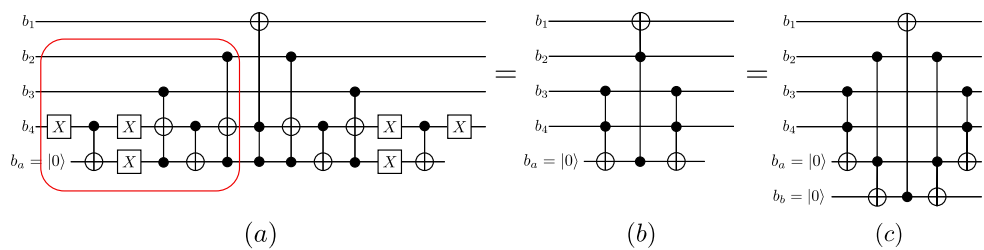}
\caption{We can simplify the circuit in (a) to the one illustrated in (b). In (c) additional  Toffoli gates are used to make a general C$^{3}$U. All the lines represent qubits.}
\label{simplify}
\end{figure}

As our scheme is a direct conversion from a ququart scheme, we have generated a gate sequence where all our Toffoli gates must act on both qubits which were formerly part of the ququart. The gates circled in fig \ref{simplify}(a) leave the ancilla qubit in $|1\rangle$ only if qubits three and four are initially in $|1\rangle$, and leave qubit four in $|1\rangle$ only if qubit two, and qubit four are in the state $|1\rangle$. The first operation is easy to create using a single Toffoli, but the second operation is non-unitary so cannot be created easily without the addition of an ancilla. However, the second expression also has a level of redundancy and can be replaced by a Toffoli gate which flips the target qubit only if the ancilla qubit, and qubit three are in $|1\rangle$. The result is the circuit shown in fig \ref{simplify}(b), which is directly equivalent to previous results in Nielsen and Chuang \cite{Nielsen2000}. In fig \ref{simplify}(c) we make a small adaptation, adding an additional Toffoli gate, CNOT gate, and ancilla qubit. 

From simple counting arguments we find that the total number of Toffoli gates required to implement this sequence for a gate of the form C$^{n}$X is $2n-3$ When we limit our gate set to consist of CNOT and local operations we need $14n-13$ operations, when we have the ability to perform the CV gate, we only need $8n-11$ operations to generate our C$^{n}$X gate compared to the $12n-22$ required by Miller \emph{et al.}~\cite{Miller2009,Miller2011}. We can therefore clearly see that a simplification of the Lanyon scheme \cite{Lanyon2008} is equivalent to the scheme in Nielsen and Chuang \cite{Nielsen2000} and that scheme is more efficient than other optimizations for creating large controlled gates. 

\begin{figure}
\includegraphics[width=8.4cm]{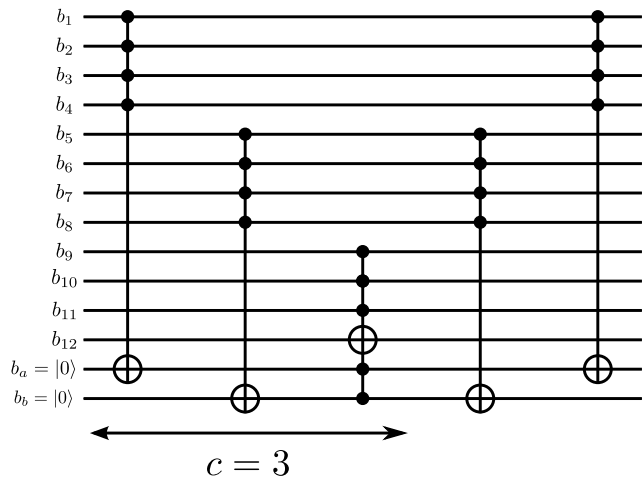}
\caption{Creating a large Toffoli gate from several smaller Toffoli gates, this uses two ancilla qubits which start in the state $|0\rangle$ and makes a Toffoli which is controlled on 11 qubits. Additional ancilla qubits will be required to create the shown gates.} 
\label{cycle}
\end{figure}

However, the scheme provided in Nielsen and Chuang \cite{Nielsen2000} requires a fixed number of ancilla, while more recent research \cite{Miller2009,Miller2011} provides techniques for implementing large controlled gates using any number of ancilla qubits. We therefore want to look at how to minimize the number of ancilla qubits required to generate a large controlled unitary using initialized ancilla. To do this we use the identity in Barenco \emph{et al.}~\cite{Barenco1995} which combines two copies of C$^{f}$X and two copies of C$^{m}$X to create C$^{f+m}$X using only a single ancilla qubit. In fig \ref{cycle} we show how we can use this identity to reduce the number of ancilla qubits used to generate a large controlled unitary. Generating one multiple qubit controlled gate can be considered one cycle. Our procedure consists of $2c-1$ cycles, where the first $c$ cycles are used to flip our target only if all the control qubits are in $|1\rangle$ and the other $c-1$ cycles are used to return all the ancilla qubits to $|0\rangle$

The first $c$ cycles of our process flip the target qubit, but also return all but $c$ of our qubits to their initial state. This set of cycles therefore requires a total of $2n-2-c$ Toffoli gates. The average number of Toffoli gates per cycle is therefore
\begin{equation}
N_{c}(n,c) = \frac{2n-2-c}{c} 
\end{equation}
Given we require $2c-1$ cycles the total number of Toffoli gates for creating an $n$ qubit controlled unitary using $2c-1$ cycles, $N_{t}(n,c)$, is
\begin{equation}
N_{t}(n,c) = (2c-1) \left( \frac{2n-2-c}{c} \right) 
= \left\lfloor \frac{2n(2c-1)-c(3+2c)+2}{c} \right\rfloor
\end{equation}
We take the floor function here, because we can always chose to have the shortest cycles as the ones we repeat twice. 

We use an ancilla qubit as the target of all but one of our multiple qubit controlled gates, therefore we need $c-1$ ancilla qubits to act as `cycle' qubits, which are not reused between cycles. Process ancilla qubits will be used to create the multiple control gates, these will be reused in each cycle. We need $n-1$ Toffoli gates to flip our target qubit, $c$ of these will have a cycle ancilla as a target therefore $n-1-c$ will need a process ancilla as a target. The Toffoli gates are equally divided between the cycles therefore the total number of ancilla qubits required is given by
\begin{equation}
N_{a}(n,c) =  \left\lceil\frac{n-1-c}{c} \right\rceil+ c -1 = \left\lceil \frac{n-1}{c}\right\rceil + c -2
\label{Nan}
\end{equation}
We take the ceiling function here because we need enough ancilla qubits for the longest cycle. To find the minimum number of ancilla qubits we differentiate equation (\ref{Nan}) without the ceiling function, giving $c = \sqrt{n- 1}$. We take $c = \lfloor \sqrt{n-1} \rfloor$ to keep the number of operations as small as possible. Therefore we require
\begin{equation}
N_{a}(n,\lfloor \sqrt{n- 1} \rfloor) =  \left\lceil \frac{n-1}{\lfloor \sqrt{n- 1} \rfloor} \right\rceil+ \lfloor\sqrt{n- 1} \rfloor-1  \approx 2 \sqrt{n-1} -1
\end{equation}
ancilla qubits. This gives us a quadratic reduction  in the number of ancilla qubits required. However as $n$ becomes large, the number of operations required to achieve this minimum tends to
\begin{equation}
N_{t}(n, \sqrt{ n- 1}) = 4(n-\sqrt{n})
\end{equation}
meaning that almost double the number of operations are required to get this reduction in the consumption of ancilla qubits.  For general $n$ and $c = \lfloor \sqrt{n- 1} \rfloor$ the total number of operations assuming the ability to perform our controlled gate using CV is given by 

\begin{equation}
N_{g}(n, \sqrt{n- 1 }) = 4\left(4n - \left\lfloor \frac{2(n-1)}{\lfloor \sqrt{n-1} \rfloor} \right\rfloor- 2\lfloor \sqrt{n-1 } \rfloor-3 \right) + 2\lfloor \sqrt{n-1}\rfloor-1 
\end{equation}

This equation takes into account we need four operations to perform the majority of our Toffoli gates, while $2c-1$ of our Toffoli gates require five operations. When $n-N_{a} > 5$ then Miller et al.~\cite{Miller2011} require a total number of operations given by 
\begin{equation}
N_{g_\mathrm{M}} = 24n-64-12\left\lceil \frac{n-1}{\lfloor \sqrt{n-1} \rfloor} \right\rceil-12\left\lfloor\sqrt{n-1}\right\rfloor
\end{equation}
for the same number of ancilla qubits as we need \footnotemark[17]. We therefore expect to use less operations than Miller \emph{et al.}~\cite{Miller2011} provided $n>10$, however since the formula we derived from the work of Miller \emph{et al.}~is only accurate if $n-N_{a} > 5$ then our comparison is only accurate if $n>10$ so we could see an improvement for lower values of $n$.

\begin{table*}
\centering
\begin{tabular}{|c|c|c|c|c|c|c|c|c|c|c|c|c|c|c|c|}
\hline
n & 3 & 4 & 5 & 6 & 7 & 8 & 9 & 10 & 11 & 12 & 13 & 14 & 15 \\
\hline
Number ancilla &2 & 3 & 3 & 4 & 4 & 5 & 5 &5 & 6 & 6 & 6 & 7 & 7 \\
\hline
Our gate requirement & 13 & 21 & 39 & 51 & 63 & 75 & 87 & 105 & 121 & 133 & 145 & 161 & 173 \\
\hline
Miller's gate requirement & 14 & 26 & 38 & 50 & 64 & 76 & 96 & 116 & 128 & 152 & 176 & 188 & 212 \\
\hline
\end{tabular}
\caption{A comparison between the number of gates we require to make a gate of the form C$^{n}$X compared to those required by Miller et al.~\cite{Miller2011}. Both schemes use the Peres gate to break up the Toffoli operations.}
\label{table}
\end{table*}

In table \ref{table} we show that we require fewer operations than Miller \emph{et al.}~\cite{Miller2011} for the same number of ancilla qubits for all $n$ except $n=5$ and $n=6$. In general, as $n$ becomes larger the improvement we show over Miller \emph{et al.}~also becomes larger. We can also compare our results with those obtained by Barenco et al.~who use a system of only two cycles \cite{Barenco1995}. In this case we need $3(n-2)$ Toffoli gates compared to the $8(n-5)$ required by Barenco \emph{et al.} \cite{Barenco1995}. Both schemes will use the same number of ancilla qubits. This shows the significant advantage of initialising the ancilla qubits. 

In this paper we showed that the qubit equivalent of the qudit schemes proposed by Ralph \etal~\cite{Ralph2007} and Lanyon \etal~\cite{Lanyon2008} is directly equivalent to the scheme given in Nielsen and Chaung \cite{Nielsen2000}. Simple counting arguments show that this is currently the most efficient way to generate large controlled gates, although it is possible that optimization techniques used in other work \cite{Maslov2003,Maslov2005,Scott2008,Wille2008,Miller2009,Miller2011} could also be used in this scenario to get further reductions. We can reduce the number of ancilla qubits required by our system by creating several large Toffoli gates, then combine them to form one even larger Toffoli gate. This adaptation can double the number of operations but can give us a quadratic reduction in the number of ancilla qubits required. The minimum number of ancilla qubits required by our scheme to produce a $C^{n}X$ gate is given by
\begin{equation}
N_{a}(\mathrm{min}) = \left\lceil \frac{n-1}{\lfloor \sqrt{n- 1} \rfloor} \right\rceil + \lfloor \sqrt{n- 1} \rfloor -1
\end{equation}
For large $n$ this would require $4(n-\sqrt{n})$ Toffoli operations, roughly double the number needed if we use $n-2$ ancilla qubits. We therefore see a trade off between the number of ancilla qubits and the total number of operations required. When we reduce the number of ancilla qubits, we still require fewer total operations than needed by Miller except when $n=5$ or $n=6$, this comparison is shown in table \ref{table}. It is worth noting that it might be possible to obtain further improvements using the automated searching techniques provided in these papers. 

We show that using initialized ancilla it is possible to get a saving in operations over alternative techniques for both high and low number of ancilla. This work clearly shows the advantages of using initialized ancilla for creating large controlled unitaires. The initialization of ancilla qubits is essential for quantum error correction and is generally considered a relatively trivial procedure. The one disadvantage of this scheme is there is a limit to how far we can reduce the number of ancilla, and we show that the minimum number of ancilla required is roughly $2\sqrt{n-1}-1$. We hope that it would be possible to obtain further improvements in the number of operations required using the optimization techniques discussed in previous work \cite{Maslov2003,Maslov2005,Scott2008,Wille2008,Miller2009,Miller2011} .

\begin{acknowledgements} 
Katherine Brown and Jonathan Dowling are supported by the Intelligence Advanced Research Projects Activity (IARPA) via Department of Interior National Business Center contract number D11PC20168. The U.S. Government is authorized to reproduce and distribute reprints for Governmental purposes notwithstanding any copyright annotation thereon. Disclaimer: The views and conclusions contained herein are those of the authors and should not be interpreted as necessarily representing the official policies or endorsements, either expressed or implied, of IARPA, DoI/NBC, or the U.S. Government. Sabre Kais thanks NSF CCI Award CHE-1037992. Jonathan Dowling also acknowladges the NSF \& the AFOSR.
\end{acknowledgements} 

\bibliographystyle{Science}
\bibliography{bibliography}

\end{document}